# A Radiomics-Boosted Deep-Learning Model for COVID-19 and Non-COVID-19 Pneumonia Classification Using Chest X-ray Image

Zongsheng Hu[1], Zhenyu Yang[2], Kyle J. Lafata[2,3,4], Fang-Fang Yin[1,2], Chunhao Wang*[2]

[1]Medical Physics Graduate Program, Duke Kunshan University, Kunshan, Jiangsu, China 215316
[2]Department of Radiation Oncology, Duke University, Durham, NC, 27710
[3]Department of Radiology, Duke University, Durham, NC, 27710
[4]Department of Electrical and Computer Engineering, Duke University, Durham, NC, 27710

Running Head: Radiomics-boosted COVID-19 detection

*Corresponding authors:
Chunhao Wang, Ph.D.
Department of Radiation Oncology
Duke University Medical Center
Durham, NC, 27710, United States
E-mail: chunhao.wang@duke.edu

# Abstract

Purpose:

To develop a deep-learning model that integrates radiomics analysis for enhanced performance of COVID-19 and Non-COVID-19 pneumonia detection using chest X-ray image.

Methods:

Based on a pre-trained VGG-16 architecture, the deep-learning design consisted of a 5-Dense layer neural network with varying sizes, while the last three convolutional layers were set as free parameters for training. In radiomics analysis, a 2D sliding kernel was implemented to map the impulse response of radiomic features throughout the entire X-ray image; thus, each feature is rendered as a 2D map in the same dimension as X-ray image. For comparison purpose, a pilot model was trained using X-ray image only. Subsequently, 2 radiomic feature maps (RFMs) were selected based on cross-correlation analysis in reference to the pilot model saliency map results. The radiomics-boosted model was then trained based on the same deep-learning design using X-ray image plus the selected RFMs as input.

Both models were developed using 812 chest X-ray images with 262/288/262 COVID-19/Non-COVID-19 pneumonia/healthy cases, and 649/163 cases were assigned as training-validation/independent test sets. For each model, 50 versions were trained with random assignments of training/validation cases following 7:1 ratio in the training-validation set. Sensitivity, specificity, accuracy, and ROC curves together with Area-Under-the-Curve (AUC) were compared between the two models.

Results:

The performance of detecting COVID-19 vs Non-COVID-19 pneumonia in the pilot model using X-ray image only were:1) sensitivity 0.90±0.07 *vs* 0.78±0.09, 2) specificity 0.94±0.04 *vs* 0.94±0.04, 3) accuracy 0.93±0.03 *vs* 0.89±0.03, and 4) ROC AUC 0.96±0.02 *vs* 0.92±0.04. Entropy and Short-Run-Emphasize were selected based on their high correlations with pilot model saliency map results. In the radiomics-boosted model, the corresponding results demonstrated significant improvements (*p<0.05*) of COVID-19 *vs* Non-COVID-19 pneumonia detection: 1) sensitivity 0.95±0.04 *vs* 0.85±0.04, 2) specificity 0.97±0.02 *vs* 0.96±0.02, 3) accuracy 0.97±0.02 *vs* 0.93±0.02, and 4) ROC AUC 0.99±0.01 *vs* 0.97±0.02. The reduced variations suggested a superior robustness of the radiomics-boosted model.

Conclusions:

The inclusion of radiomic analysis in deep-learning model improved the performance and robustness of COVID-19/Non-COVID-19 pneumonia detection, which holds great potential for clinical applications in the COVID-19 pandemic.

# Introduction

Since its first discovery in 2019, the coronavirus disease (COVID-19) has affected more than 100 million people globally, and 2.7 million deaths related to COVID-19 were reported by April 1$^{st}$, 2021[1]. Accurate and efficient diagnosis of COVID-19 is crucial to interrupt disease transmission and to start treatments of affected individuals. Currently, reverse transcription polymerase chain reaction (RT-PCR) has been recognized as the gold standard for COVID-19 diagnosis for its high specificity[2]. While RT-PCR test may have limited sensitivity and longtime of processing (a few hours to 2 days)[3], radiographic procedures, including chest X-ray and CT exams, have been adopted clinically as alternative diagnosis tools[4]. While COVID-19 related abnormalities could be more easily found in volumetric CT images[5], planar chest X-ray has its unique advantages in COVID-19 diagnosis. Specifically, the short imaging time on a more accessible X-ray unit enables rapid COVID-19 exams, which can be critical in areas with high-volume patients and/or limited-resource medical facilities. To date, pilot studies have revealed that certain X-ray image features, including peripheral consolidations and ground-glass opacities, have been widely observed in COVID-19 infected patients[4-9]. However, the prevalent application of chest X-ray imaging in COVID-19 diagnosis is challenged by relatively limited sensitivity and specificity[5]. Additionally, radiographic exams including chest X-ray may not be optimal for radiologists' reading in the differentiation of non-COVID-19 pneumonia from COVID-19, which is important for early patient stratification that can lower COVID-19 mortality rate with more targeted treatments[10,11].

Computed aided diagnosis systems (CAD) may have the potential to solve the aforementioned problem with high throughput quantitative analysis. In the last several months under the COVID-19 pandemic, studies revealed that CAD systems outperformed radiologists in radiographic-based COVID-19 diagnosis; with CAD information as reference information, radiologist reading results could be significantly improved[11-13]. One approach for such CAD system is radiomics-based image analysis, which first extracts radiomics features as computation image biomarkers and then uses the extracted features in hand-made or machine learning classifier tasks. Although handcrafted radiomics features are commonly used in medical image analysis with possible qualitative image interpretability, the reported accuracy (75-80%) of COVID-19 diagnosis is still limited in the representative radiomics-based CAD works[14-16]. Driven by recent theoretical developments and accesses to massive computation power, deep learning has demonstrated its great potential in CAD developments. It has been reported that deep learning solution based on artificial neural network deployment could achieve high (>90%) specificities in COVID-19 diagnosis against healthy individuals[17,18]; moreover, decent specificities (>85%) of differentiating COVID-19 from non-COVID-19 pneumonia have been achieved[11,19-21]. Nevertheless, like all other deep learning applications in medical image analysis, the hyperparameters in the neural network are generated without explicit human knowledge intervention. Thus, the 'black box' nature of deep learning-based CAD inhibits their interpretability, and potential clinical applications of these CADs could be impaired by limited interpretability by clinicians.

In this work, we aim to develop a radiomics-boosted deep learning CAD design for chest X-ray based COVID-19 diagnosis. As hand-crafted radiomics and deep learning are complementary approaches to image representation, their integration may facilitate better model performance and interpretation[22]. Based on a custom-trained neural network, our model classifies COVID-19, non-COVID-19 pneumonia, and healthy individuals. An innovative implementation of radiomics analysis was included to analyze neural network parameters, and the analysis results were used to incorporate novel radiomics results in image rendering as neural network inputs. Thus, the achieved radiomics-boost deep learning can be better interpreted by qualitative image descriptions within in current clinician knowledge domain.

## Materials and Methods

*Image Dataset*

In this IRB-waived retrospective study, a total of 812 chest X-ray images were collected from three public databases[23-25], including 262/288/262 images of COVID-19/Non-COVID-19 pneumonia/healthy individuals, respectively. The image numbers from the three categories were approximately the same for eliminating categorical bias during deep learning training. All collected images were verified by experienced medical physicists with proper lung X-ray display settings and no overlaid image reading annotations. To unify image data size, all images were resized to a 256x256 matrix grid size using b-spine interpolation and were normalized to 256 gray levels. 649 and 163 images (8:2) were assigned for the model training set and the independent test set, respectively.

*Neural Network Architecture*

Figure 1 illustrates the deep learning neural network architecture of this work. This design was mainly constructed based on a pre-trained VGG-16 architecture[26], which consists of two parts: the 1$^{st}$ part is convolutional base with 5 convolutional blocks. Each convolutional block is stacked by 2 or 3 convolutional layers and a max pooling layer. In each convolutional layer, the filter size is 3×3 with padding and stride of 1. Max-pooling is performed over a 2x2-pixel window with a stride of 2; the 2$^{nd}$ part is the Dense part,which is the stack of Dense layers. Depending on specific classification tasks, the number and size of Dense layers can be customized. In this work, the self-defined Dense classifier connects with convolutional base and consists of five Dense layers with the size of 1024, 1024, 512, 256, and 3, respectively. The input of neural network is a three-channel image with a 256×256×3 shape size, while the output is one of the three categorical binary label vectors, i.e., [1,0,0], [0,1,0], and [0,0,1], which correspond to COVID-19, Non-COVID-19 pneumonia, and healthy results, respectively. To deal with relatively small data size in this work, the convolutional base loaded the weights that were pre-trained on ImageNet as a transfer learning scheme[27]. In addition, in order to make the model more relevant for the problem at hand, fine-tune technique was used with the last three convolutional layers (marked * in Figure 1) being set as free parameters for task-specific training. To avoid the occurrence of overfitting, a dropout layer was added between the first two Dense layers with a dropout possibility of 0.5, and soft-max activation was used in the output layer.

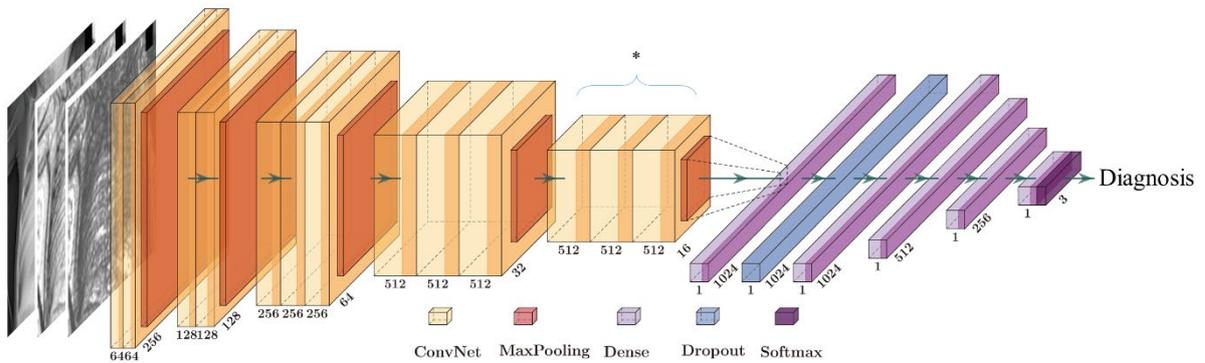

*Figure 1. Deep learning network diagram.*

*Radiomic Feature Map Extraction*

Classic radiomics analysis calculates radiomic features as scalar values from a pre-defined region-of-interest (ROI) in image space. While this approach has been widely adopted to capture the overall textures in ROI, it cannot capture the anatomy-driven subtle texture variations within the ROI. As such, we implement a radiomic feature map calculation workflow[28], which is summarized in Figure 2. For RFM generation, a 2D kernel (13x13 matrix size) was adopted to form a region-of-interest (ROI), and 37 radiomic features were extracted as a 1x37 vector within this ROI following classic GLCOM[29] (21 features) and GLRLM[30] (16 features) feature extraction methods using 32 grey levels. For each feature, the calculated feature value was assigned as the pixel value that centered at the ROI. By moving this 2D kernel across X-ray image as a sliding window operation, 37 feature maps were formed in the same dimension as original X-ray images. All radiomic anlaysis was done using custom code that was benchmarked with digital phantoms[31] and complies with the imaging biomarker standardization initiative[32].

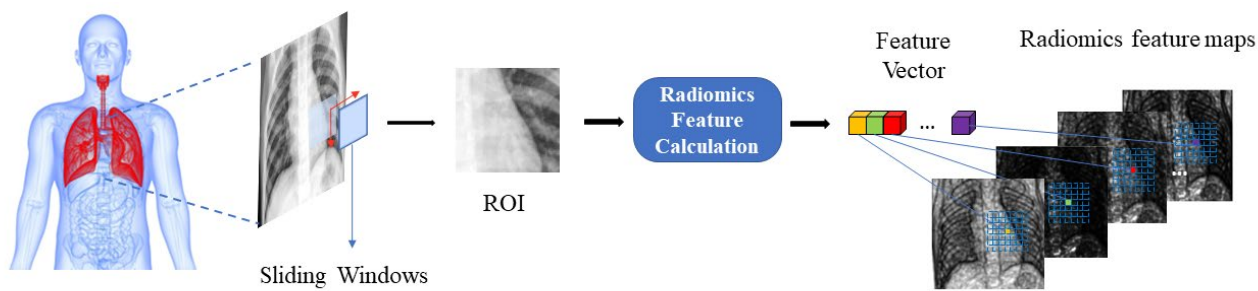

*Figure 2. A workflow summary of radiomic feature map (FM) calculation in this work.*

*Model Training and Evaluation*

Two deep learning models were developed following the neural network architecture in Figure 1. In the 1st pilot model, the X-ray image is the sole model input. To accommodate the VGG-16 data shape, the grayscale X-ray images were broadcast to three channels as a network input variable. This model serves as the benchmarking deep learning model in this work. In the 2nd radiomics-boosted deep learning model, the grayscale X-ray image and two derived RFMs were stacked as the 3-channel neural network input variable. These two RFMs were selected based on the analysis of the 1st model's saliency map (SM), which indicates how important each pixel is with respect to the final classification results of the neural network in the benchmarking model. It is calculated as the absolute gradient of class activation, which is defined as the dot product of prediction output and target divided by the input image[33]: a pixel with a higher value in SM indicates higher importance of that pixel in neural network's attention for diagnosis. The two RFMs (one in each feature category) with the highest average cross-correlation (CC) values against the SM results in training data were selected. This action amplifies certain pixels (and regions) with potentially high importance of disease diagnosis in the image space, which could improve the overall diagnostic accuracy of the proposed radiomics-boosted model in comparison with the 1$^{st}$ pilot model.

To investigate the robustness of the two developed models, 50 versions of each model design were trained using the training data set (649 images). In each version, the training and validation samples were randomly selected following a 7:1 ratio. During deep learning training, the loss function was categorical cross-entropy, and Adam optimizer was selected. The training time was about 15 minutes in a TensorFlow environment using an Nvidia$^{TM}$ (Santa Clara, CA) Tesla V100 graphic card. For model evaluation, sensitivity, specificity, accuracy, and ROC area under the curve (AUC) results from both models were analyzed. Statistical significance of comparison was determined by Wilcoxon signed rank tests at level 0.05

# Results

Based on the saliency map analysis of the 1st pilot model, two RFMs, GLCOM Entropy (cc=0.33) and GLRLM Short Run Emphasis (SRE) (cc=0.31) were found to have the highest CC among their categories, respectively. The stack of two selected RFMs and original X-ray was used for radiomics-boosted deep learning model inputs.

Figure *3* shows the identified RFMs from three example patients. As illustrated, the RFMs render more tissue textural variations in both lung and other soft tissue regions.

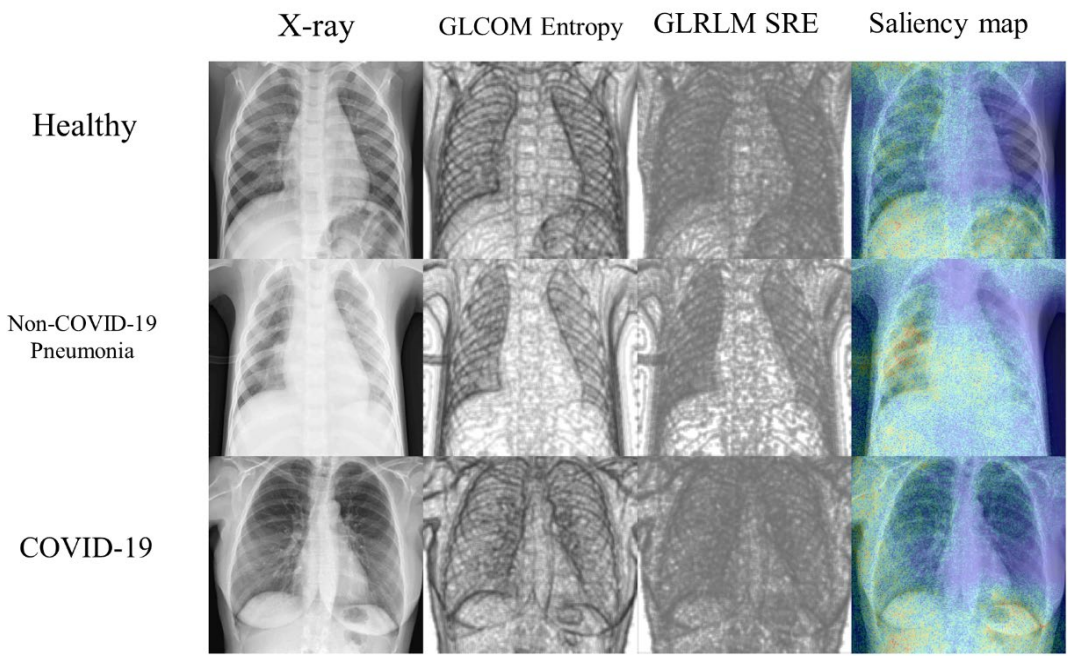

Figure *3. Image comparisons from 3 example cases. The GLRLM SRE RFMs and saliency map (overlaid with X-ray image) are illustrated in 0.3 power scale.*

Figure *4* summarizes the ROC analysis results for two models. The blue and red solid lines represent the average ROC results of 50 versions of two deep learning models (X-ray only vs. X-ray + RFM), and the colored bands represent the model performance variation of 50 versions as ±1 standard

deviation. As demonstrated, the average result of the radiomics-boost deep learning model (X-ray + RFM) outperformed the pilot model using X-ray only. The largest performance improvement is observed in Non-COVID-19 pneumonia diagnosis. Additionally, the radiomics-boost deep learning model has narrower ROC band width, which suggests the enhanced robustness of its design under different data sample uses.

Table 1 summarizes the quantitative comparisons of sensitivity, specificity, accuracy, and ROC AUC between two tested DL models. In general, the radiomics-boost deep learning model achieved statistical improvements in all parameters with $p < 0.05$. The largest improvements were observed in non-COVID-19 pneumonia diagnosis. Additionally, the reduced standard deviations of the reported statistics indicated the enhanced robustness of radiomics-boost deep learning design. These quantitative results highlight the superiority of the proposed radiomics-boost deep learning model. It worth mentioning that the developed radiomics-boost deep learning model achieved the best performance in COVID-19 diagnosis. For COVID-19 diagnosis, the developed radiomics-boost deep learning model achieved higher sensitivity (94.9%), specificity (97.3%), accuracy (96.5%), and AUC(99.3%) compared to the pilot model. The standard variation of 50 versions is less than 3%, reaching a high level of robustness. The performance demonstrates the potential of enhancing the accuracy and efficiency of COVID-19 diagnosis in clinical application by incorporating radiomics information in DL model.

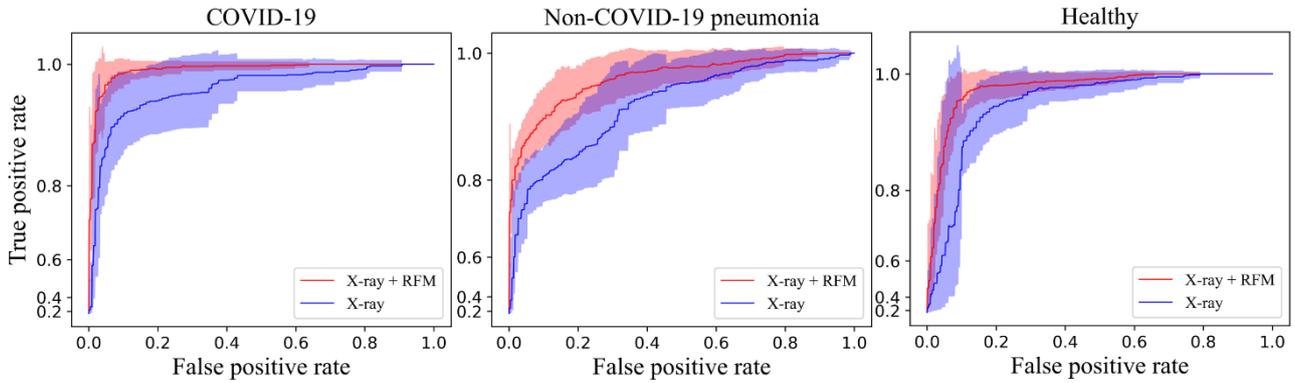

*Figure 4. The ROC results from 2 different deep learning model design. 0.3 power scale was used in y axis to highlight the difference.*

|  | Healthy | | Non-COVID-19 pneumonia | | COVID-19 | |
| --- | --- | --- | --- | --- | --- | --- |
|  | X-ray | X-ray+RFM | X-ray | X-ray+RFM | X-ray | X-ray+RFM |
| Sensitivity | 0.854±0.065 | 0.922±0.059 | 0.780±0.092 | 0.857±0.0361 | 0.903±0.071 | 0.949±0.036 |
| Specificity | 0.918±0.044 | 0.938±0.022 | 0.941±0.041 | 0.963±0.023 | 0.940±0.037 | 0.973±0.020 |
| Accuracy | 0.895±0.029 | 0.933±0.023 | 0.892±0.029 | 0.931±0.016 | 0.927±0.028 | 0.965±0.016 |
| AUC | 0.948±0.027 | 0.979±0.012 | 0.918±0.043 | 0.969±0.017 | 0.963±0.023 | 0.993±0.006 |

*Table 1. Sensitivity, specificity, accuracy, and ROC AUC results summary. Mean value and standard deviation of 50 trained versions are reported. All comparisons between pilot model (X-ray) and radiomics-boosted model (X-ray + RFM)   were statistically significant*

The SMs results of the radiomics-boosted deep learning model are illustrated in the last column of Figure 3. The pixel values in SM can be interpreted as the attention of the deep learning model. As seen, the attention patterns, i.e., colored hot regions distribution in SMs, were different in different patient cohorts: the model may have more attention to lateral lung regions for COVID-19 detection, while it may have more attention to the mediastinum regions for non-COVID-19 pneumonia detection.

In order to quantitatively analyze the attention patterns across different patient cohorts, we calculated the CC matrix of SMs in the test set in Figure 5, which includes all the CCs between SMs of 2 samples from the test set. As summarized, CCs within each cohort were relatively higher than those calculated across different cohorts. This result suggests that the developed deep learning model captured cohort-specific features for the classification task. Additionally, the mean CC result of COVID-19 *vs* Non-COVID-19 pneumonia (0.12) cohort was slightly higher than the result of COVID-19 *vs* Healthy cohorts (0.07) and Non-COVID-19 pneumonia *vs* Healthy cohorts (0.09). This observed COVID-19/Non-COVID-19 pneumonia similarity supports the clinical reports of challenges in COVID-19/Non-COVID-19 pneumonia differentiation[10,11].

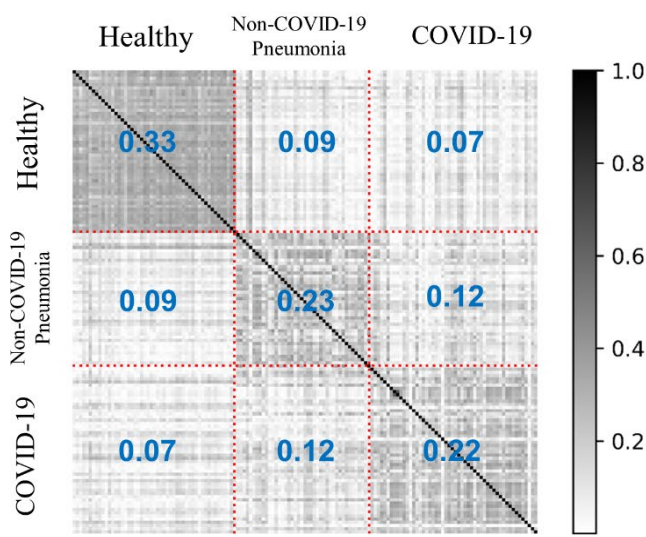

*Figure 5. The cross-correlation matrix of SMs of 2nd model on test set. The x and y axis represents the sample ID in test set, sorting with the order of healthy/Non-COVID-19 pneumonia/COVID-19 cohorts.*

## Discussion

To our best knowledge, this work is the first of its kind for combining radiomic analysis and deep neural network implementation. The results of this work demonstrated that the inclusion of radiomic feature maps, as a new form of handcrafted imaging biomarker rendering, can improve deep learning-based COVID-19 detection. With the aid of RFMs, we achieve higher model performance (a sensitivity of 92.2%/85.7%/94.9%, a specificity of 93.8%/96.3%/97.3%, and an accuracy of 93.3%/93.1%/96.5% for Healthy/non-COVID-19 pneumonia/COVID-19) with a smaller (812 patients in total) than reported work. For example, Zhang et al.[11] achieved a sensitivity of 88% and a specificity of 79% in COVID-19/non-COVID-19 pneumonia diagnosis with a dataset of 2060 patients. Nishio et al.[19] achieved an accuracy of 83.7% for 3 categories of classification (Healthy/non-COVID-19 pneumonia/COVID-19) using the VGG-16 model trained on 1248 images. Tulin et al.[21]. achieved an accuracy of 87.0% for tri-class classification (Healthy/non-COVID-19 pneumonia/COVID-19) using the Darknet-19. In this work, we studied a total of 812 chest X-ray images from 3 public datasets[23-25], which were not curated by the typical medical image study protocols. As a result, proper image processing is necessary for streamlined deep learning implementation. In particular, we resized all images to 256x256 grid size and normalized all images to 256 gray levels as uint8 format. These operations are standard in digital image processing which will facilitate data reproducibility of this work. In addition, the robustness of the developed model was systematically analyzed. For each model design, we trained 50 versions of models using randomly selected training and validation samples following a ratio of 7:1. The small standard deviation (< 0.06) of selected metrics and ROC results revealed the enhanced robustness of the developed model, which further demonstrates the potential of the radiomics-boosted deep learning design in clinical situations using different X-ray image data sources.

The deep learning implementation in this work adopted a VGG-16 based transfer learning scheme. VGG-16 was selected for the following reason. First, it has been widely studied for medical image analysis tasks as transfer learning scheme. Second, in comparison with other prevalent candidates (such as ResNet block-based architectures), VGG-16 possesses a lightweight architecture design and thus leads to reduced calculation workload for network training. Finally, previous studies[19] reported that VGG-16 achieved the highest accuracy in COVID-19 diagnosis tasks in comparison with several other pre-trained deep neural network architectures.

The inclusion of radiomic feature maps (RFMs) is another key technical innovation. Instead of calculating radiomic features as scalar values from selected volumes in image space, RFMs capture the anatomy-driven subtle texture variations within ROIs. It has been demonstrated that radiomics are associated with pulmonary function[34] and lung ventilation measurements[28]; as such, the potential functional information in RFMs contributes to the enhanced COVID-19 diagnosis accuracy. The selection of two RFMs from 37 RFMs is more than a trivial task: while direct comparisons of all possible RFM combinations are feasible, it requires high computational cost without the potential of transferring this technique to other clinical applications. Driven by the hypothesis that certain RFMs can be related to neural network hyperparameters, we selected RFMs based on similarity metrics between RFMs and neural network saliency maps (SMs), which measures the attention pattern of network implementation. Such pattern can be used as an auxiliary tool for radiologists in image reading, i.e., highlights specific regions as visual clues for human reading. This potential human-aid tool can be important in accurate COVID-19/ NON-COVID-19 pneumonia differentiation, which can be a challenging task for radiologist using chest X-ray images without volumetric information[11]. RFMs with higher similarities to SMs could emphasize regional information to enhance neural network attention, which increases synchronously with SM results. In current clinical practice, mutual information (MI) and cross-correlation (CC) are two common image similarity metrics for image registration and classification. Figure 6 summarizes the top 8 RFM candidates in GLCOM and GLRLM feature categories with normalized MI and CC results. As seen, while RFMs were sorted by CC values, most of them had close normalized MI results, which might not be directly related to image difference appearance; Given these considerations, two RFMs, GLCOM Entropy (CC = 0.33) and GLRLM Short Run Emphasis (SRE) (CC = 0.31) from two feature categories were selected based on CC analysis. It would be of interest to investigate other RFM selection mechanisms that can be complementary to the current design. Additionally, the combination of radiomics feature map analysis and other deep neural network candidates, such as EfficientNet and VGG-19, could be studied following the current methodology in future works.

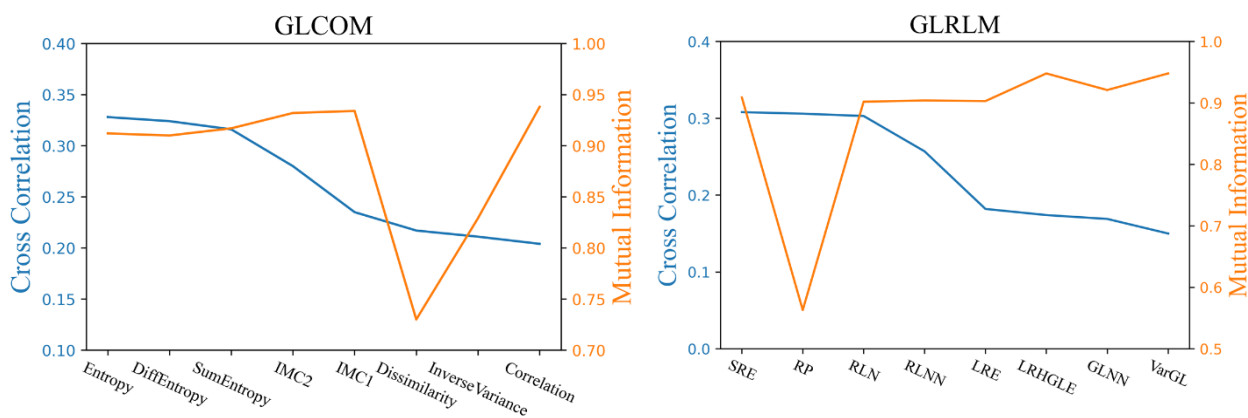

*Figure 6 Normalized MI and CC results of top 8 RFM candidates ranked by CC results in GLCOM and GLRLM feature categories*

The presented design of combining radiomics analysis and deep neural network implementation may create a new paradigm of CAD system. The implemented RFM calculation workflow may also enhance neural network performance in other tasks, particularly those where multi-channel imaging data are required as input[35]. Additionally, the proposed method provides a radiomics perspective of deep learning interpretability. The hyperparameters in the neural network are trained without explicit human knowledge intervention and thus are hard to interpret by empirical knowledge. For deep learning-based CAD systems, the 'black box' nature impaired the clinical deployments of such systems without clinicians' confidence. As a step towards deep learning interpretability, we investigated neural network attention information using a radiomics-based analysis. Radiomics has been widely studied as computational imaging biomarkers for disease detection and outcome monitoring[36], and it has been demonstated that radiomic feature feature spaces can be mathematically decomposed to provide interpration[37]. Following saliency map analysis approach in this work, additional parameters can be used to enhance deep learning interpretability, such as histology samples image from biopsy and anatomy contours from radiation therapy. These directions will be studied in future works when appropriate datasets become available.

# Conclusion

In this study, we proposed a radiomics-boosted deep learning design for X-ray based COVID-19 diagnosis and non-COVID-19 pneumonia diagnosis. An innovative radiomic feature map calculation workflow was implemented to generate inputs of a VGG-16 based neural network architecture. Results showed that the proposed radiomics-boosted deep learning design improved the performance and robustness of COVID-19 and Non-COVID-19 pneumonia detection in concurrence with a radiomics viewpoint of deep learning interpretation. It holds great potential for clinical applications for COVID-19 diagnosis and for other diseases.

# Figure Captions

Figure 1. Deep learning network diagram.

Figure 2. A workflow summary of radiomic feature map (FM) calculation in this work.

Figure 3. Image comparisons from 3 example cases. The GLRLM SRE RFMs and saliency map (overlaid with X-ray image) are illustrated in 0.3 power scale.

Figure 4. The ROC results from 2 different deep learning model design. 0.3 power scale was used in y axis to highlight the difference.

Figure 5. The cross-correlation matrix of SMs of 2nd model on test set. The x and y axis represents the sample ID in test set, sorting with the order of healthy/Non-COVID-19 pneumonia/COVID-19 cohorts.

Figure 6 Normalized MI and CC results of top 8 RFM candidates ranked by CC results in GLCOM and GLRLM feature categories